\documentclass[final,5p,times,twocolumn]{elsarticle}

\usepackage{amssymb}
\usepackage{mathtools}

\journal{Radiation Measurements}

\begin{document}

\begin{frontmatter}



\title{An easy tool for the Monte Carlo simulation of the passage of photons and electrons through matter}


\author{Víctor Moya}
\author{Jaime Rosado}
\author{Fernando Arqueros}

\affiliation{organization={IPARCOS-UCM, Instituto de Física de Partículas y del Cosmos and EMFTEL Department, Universidad Complutense de Madrid},%Department and Organization
            %addressline={Address One}, 
            city={Madrid},
            postcode={E-28040}, 
            %state={State One},
            country={Spain}}

\begin{abstract}
A simple Monte Carlo (MC) algorithm for the simulation of the passage of low-energy gamma rays and electrons through any material medium is presented. The algorithm includes several approximations that accelerate the simulation while maintaining reasonably accurate results.
%%JR: limitaciones del código
Notably, pair production and Bremsstrahlung are ignored, which limits the applicability of the algorithm to low energies ($\lesssim 5$ MeV, depending on the medium).
Systematic comparisons for both photons and electrons have been made against the MC code PENELOPE and experimental data to validate the algorithm, showing deviations in the deposited energy smaller than or around 10\% in the energy interval of $0.1 - 5$~MeV in light media. The simulation is also valid for heavy media, but with less accuracy 
%%JR: limitaciones del código
as a consequence of
%%FAM 
%%of ignoring 
the abovementioned approximations. Also X-ray fluorescence is ignored leading to some limitations for photons with energies slightly higher than the K-shell.  
%%some effects, such as X-ray fluorescence and Bremsstrahlung.
%%
The algorithm has been implemented in an open-source Python package called LegPy, which provides an easy-to-use framework for rapid MC simulations aiming to be useful for applications that do not require the level of detail of available well-established MC programs.
\end{abstract}




\end{frontmatter}


\section{Introduction}
\label{sec:intro}

The study of the interaction of ionizing radiation with matter is of utmost importance in a wide range of applications. Monte Carlo (MC) simulations are extensively used to this purpose, and several excellent MC programs are available, such as EGS4 \cite{egs4}, PENELOPE \cite{penelope}, GEANT4 \cite{geant4} and MCNP \cite{mcnp}, to name a few. These programs provide frameworks for detailed simulations of any case of study, including complex geometrical forms and all the physical processes that ionizing particles may undergo in a broad energy range. Moreover, these programs typically offer multiple options for different physical models and corrections due to the specific features of the atomic composition of the media. Unfortunately, such accuracy comes at the cost of increasing technical complexity, which may require users to acquire expertise to obtain meaningful results. Additionally, the more detailed a simulation is, the more time and computing resources it demands, which could limit the use of these programs in some applications.
As an alternative, analytical calculations or simple models can be used to obtain estimates of the desired results. However, this approach generally lacks the accuracy and level of detail provided by MC simulations.

In this paper we present a simplified MC algorithm for the simulation of the passage of low-energy
%%JR: rango de energía
(i.e., from about 1 keV up to a few MeV)
gamma rays and electrons through any material medium. The algorithm is based on several approximations that enhance simulation speed and simplicity while maintaining reasonably accurate results. In particular, we developed a very simple model for electron transportation that accelerates the simulations significantly. In this work, we analyzed the range of validity and assessed the impact of these approximations by comparing them against both experimental data and results from well-established MC programs, especially PENELOPE \cite{penelope}. We implemented the algorithm in a Python package called LegPy, released under an open-source license \cite{legpy}. This tool aims to provide
%%JR: aplicaciones
a generic and easy-to-use framework
for rapid simulations in applications where minor details are not necessary.

The paper is structured as follows. In section \ref{sec:algorithm}, the physical approximations employed in our MC algorithm and the main features of the package LegPy are described. The validation of the algorithm is presented in section \ref{sec:validation}. Lastly, in section \ref{sec:conclusions}, the conclusions drawn, the potential uses of our algorithm, and the improvement plans are discussed.

\section{The algorithm}
\label{sec:algorithm}

Our Monte Carlo algorithm was conceived to provide accurate enough results for a wide range of situations using a small amount of input data and computing resources. Under this approach, we neglected several effects on the transportation of photons and electrons that are only expected to be relevant at either high energy or very small scale. For the transportation of photons, pair production (energy threshold of 1.02 MeV) was ignored. Besides, several simplifications were made in photoelectric absorption, coherent scattering, and incoherent scattering. For the transportation of electrons, we developed a novel method to account for multiple scattering and collisional energy loss in a very simple and fast way. Bremsstrahlung was also ignored because it is only significant at high energy and heavy media.

%%JR: point sources/beams
%We focused on situations where a beam of photons or electrons interacts with an object made of one or more homogeneous materials. The algorithm was designed to track all the individual particles inside this object
We assume a single target object made of one or more homogeneous materials and a beam of photons or electrons generated either inside or outside this object. Only a few simple geometries for both objects and particle beams are considered to reduce the calculations and the simulation settings. All the individual particles interacting with the object are tracked, but only keeping
information on the energy deposit in a voxel-based scheme. Histograms of other relevant parameters of the beam particles (e.g., the angle and energy of escaping particles, the absorbed energy, and the maximum depth of electrons) can also be computed.

Next, the approximations made in the transportation of photons and electrons are described. In the last subsection, we briefly report on the LegPy package \cite{legpy}
%%JR: detalles
and its simulation options.

\subsection{Photon transportation}
\label{ssec:photon}

The algorithm transports photons in random steps. The distance a photon travels before it undergoes its next interaction is randomly obtained from its mean free path 
%%FAM no estaba definida 
(mfp), 
which is calculated from the total attenuation coefficient of the medium at the photon energy. The photon is transported this distance in the direction of its momentum vector to the next interaction point as long as it is inside the same medium, otherwise, the photon escapes the object or the medium changes. In the latter case, the photon is transported to the boundary between the two media to take another step in the new medium.
%%JR: detalles
If the photon is below the specified energy cut after an step, it is absorbed.

When a photon interacts, the type of interaction (i.e., photoelectric absorption, coherent and incoherent scattering) is determined at random from the relative attenuation coefficients of the different processes. Both total and relative attenuation coefficients
%%JR: interpolación
%are taken from the NIST Standard Reference Database XCOM \cite{nist_ph}.
are obtained by interpolation from tables taken from the NIST Standard Reference Database XCOM \cite{nist_ph}, which provides data down to 1 keV.
If photoelectric absorption takes place, an electron is emitted with a kinetic energy and propagation direction equal to those of the absorbed photon. In the case of incoherent scattering, the momentum vector and energy of both the photon and the electron are sampled from the differential cross section, given by the Klein-Nishina formula \cite{evans}, and the energy-momentum conservation laws. If the photon undergoes a coherent scattering, the Thompson scattering law \cite{evans} is used to randomly deviate its track.
%%JR: sampling y detalles
In our present implementation of the algorithm, a simple rejection sampling is used in both incoherent and coherent scattering, since these processes were checked to not contribute significantly to the computing time in most cases.

As already mentioned, pair production is not included in our simplified algorithm. Nevertheless, the attenuation coefficient corresponding to this process
%%JR: detalles
(also taken from the NIST database \cite{nist_ph})
is added to that of photoelectric absorption so that the attenuation of a photon beam is properly simulated
%%FAM
even for energies over the pair-production threshold.
Furthermore, the atomic effects on the above processes are ignored. In particular, the fluorescence subsequent to photoelectric absorption or incoherent scattering, i.e., the emission of X-rays by the excited atom, is not included. This is one of the main limitations of our approximation, because X-rays spread the energy deposit at larger distances than photoelectrons do. Neglecting this effect has a significant impact for heavy elements, as will be discussed later. However, our approximation is accurate enough for light elements.

\subsection{Electron transportation}
\label{ssec:electron}

The electron transportation is based on the continuous slowing down approximation (CSDA), that is, the rate of energy loss of an electron along its track is assumed to be determined by the total stopping power neglecting fluctuations. Therefore, the total path traveled by an electron is assumed to be equal to the CSDA range in the medium. Both the total stopping power and the CSDA range are
%%JR: interpolación
%taken from the NIST Standard Reference Database ESTAR \cite{nist_estar}.
obtained by interpolation from tables taken from the NIST Standard Reference Database ESTAR \cite{nist_estar}.

%%JR: detalles
%The electron path is divided into a number of steps with length chosen adequately to the desired precision, e.g., equal to the voxel size used in the simulation
The electron path is divided into a number of steps. The default option is a constant step length equal to the voxel size used to describe the object, which determines the precision of the calculations. As an alternative, the step length can be computed so that the energy loss is a constant fraction of the electron energy in each step.
Since energy-loss fluctuations are neglected, the simulation uses precomputed tables with the electron energies at the endpoints of all the steps in each constituent material of the object. When an electron is generated with a given initial energy in a medium, the energy loss and distance traveled in its first step are obtained by interpolation from the corresponding table. All the subsequent steps follow this table until
%%JR: detalles
%the electron stops
the electron energy is below the specified energy cut, it escapes the object
or the medium changes. In the latter case, the last step in the first medium is shortened to end at the boundary between the two media, and the energy loss is calculated accordingly. Then, the electron continues to be transported in the second media (if any) as if it had been generated at that point.

In each step, the end position of the electron is determined by its starting position, its propagation direction, and the step length, i.e., the electron is assumed to travel in a straight line ignoring the lateral displacement due to multiple scattering. On the other hand, after the step is taken, the electron propagation direction is randomly deviated. The axial angle of the deviation is a random number in the $0-2\pi$ interval and the scattering angle $\theta$ is sampled from a Gaussian distribution \cite{theta_0}
\begin{equation} \label{eq:gauss}
P(\theta) = \frac{1}{\sqrt{2\pi}\theta_0} \exp\left({-\frac{\theta^2}{2\theta_0^2}}\right)\,,
\end{equation}
where the average scattering angle $\theta_0$ is given by
\begin{equation} \label{eq:theta_0}
\theta_0 = \frac{E_0}{\beta cp} \sqrt{\frac{s}{X_0}} \left(1+0.038\ln\frac{s}{X_0 \beta^2} \right)\,.
\end{equation}
Here, $s$ is the step length, $X_0$ is the radiation length of the medium, $\beta$ is the ratio of the electron velocity and the speed of light $c$, $p$ is the electron momentum and $E_0$ is a model parameter. In the Gaussian approximation described in \cite{theta_0}, $E_0$ takes the value 13.6~MeV. However, we modified this parameter to compensate for the various simplifications made in electron transportation. First, we searched the values of $E_0$ that make our algorithm reproduce approximately both the energy deposit distribution and the backscattering factor obtained by PENELOPE \cite{penelope} for the energy range $0.1-5.0$~MeV and a number of media. Then, from these fitted values, we obtained the following parameterization of $E_0$ in MeV:
\begin{equation}
\label{eq:E0}
\begin{split}
      E_0 = &13.6\,(1.56+0.130x) \\
          &\left[ 1 - (0.0471-0.0182\ln\,x)\ln\,E \right]  
\end{split}
\end{equation}
%where $x=X_0^{-0.5}$ for $X_0$ expressed in cm and $E$ in MeV.
where $x=1/\sqrt{X_0}$ for $X_0$ expressed in cm and $E$ in MeV.

The necessary $\theta_0$ values are calculated at the beginning of the simulation and stored in the table of steps for each medium. The results obtained with this simple model of multiple scattering were checked to be almost independent of the step length as long as it is smaller than $10\%$ of the CSDA range.

As a further simplification, half of the energy lost by an electron in a step is deposited in the starting voxel and the other half in the end one. To avoid discontinuities in the energy deposit distribution,
%%JR: detalles
%the step length should be smaller than the voxel size.
the step length should not be larger than the voxel size.
Besides, when an electron reaches the boundary between two media, the energy loss in that step is assumed to be fully deposited in the starting voxel to prevent artifacts just at the boundary. In our approximation, no secondary particles are simulated. In particular, Bremsstrahlung is ignored, which limits the applicability of our algorithm to moderate electron energies
%%JR: limitaciones
(i.e., $<5$ MeV for heavy materials).

\begin{figure*}[t]
\centering
\includegraphics[width=\linewidth]{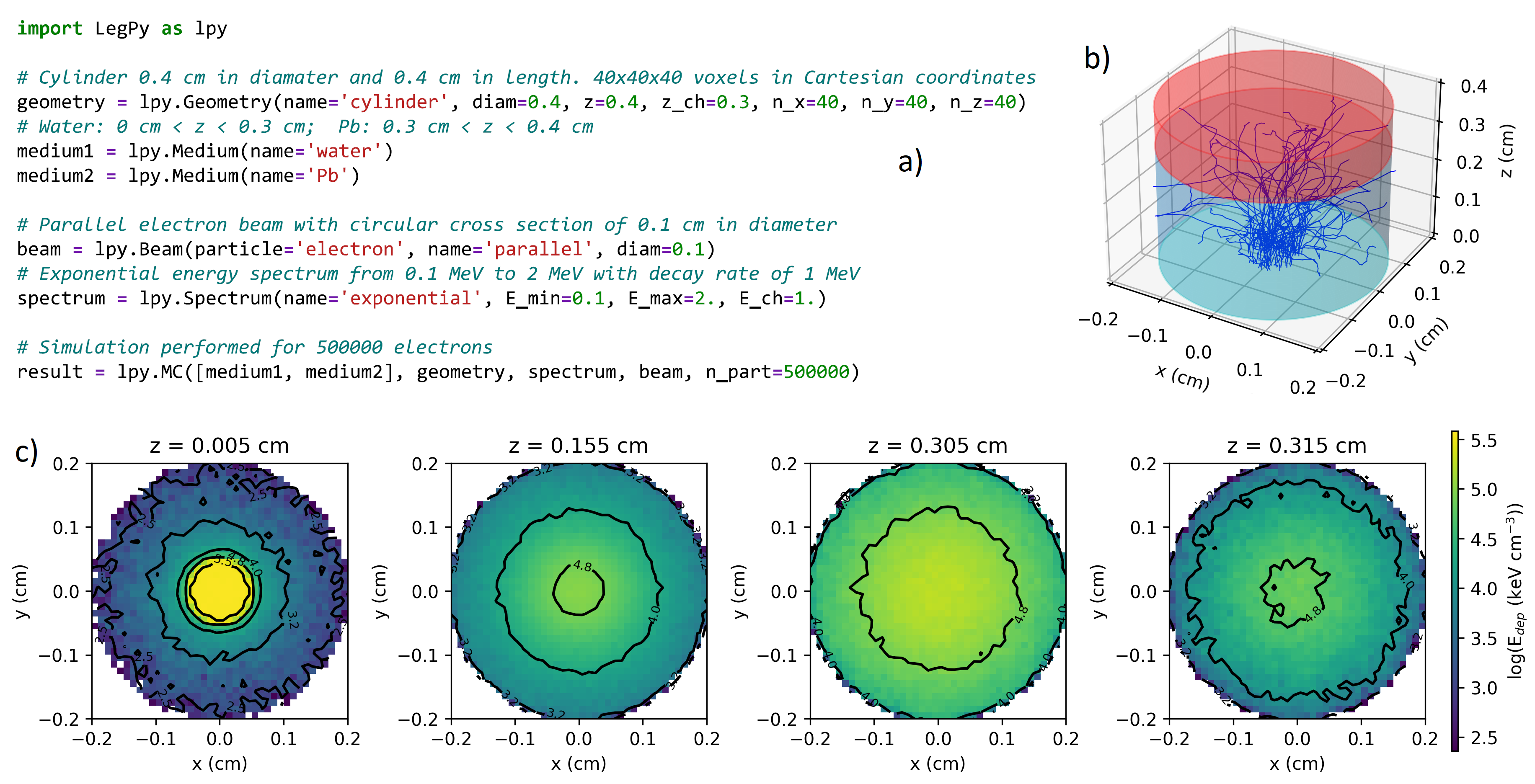}
\caption{Example of the use of LegPy. a) Sample of code that sets up and executes a simulation of a parallel beam of electrons that traverses a coaxial cylinder of water and lead. b) Tracks of 50 electrons. c) 2D distribution of deposited energy at several depths.}
\label{fig:legpy}
\end{figure*}

The use of all these approximations speeds up the simulation considerably while a reasonable accuracy is achieved
%%FAM
for many applications, 
as will be shown in section \ref{sec:validation}. The main limitation is the underestimation of the maximum range of electrons because their total path length is imposed to equal the CSDA range.

\subsection{LegPy framework}
\label{ssec:legpy}

LegPy stands for ``low energy gamma-ray simulation in Python'', but it allows the simulation of both photons and electrons using the approximations described above. The code is organized as a Python package containing classes to define the basic ingredients of a simulation: the beam geometry, the energy spectrum and type of the incident particles, the object geometry, the medium (or media) the object is made of, and the simulation settings. LegPy is designed to be used in Google Colab \cite{gcolab} or Jupyter Notebook, which are interactive environments that improve user experience. Several notebook examples are included in the last release of this software package.

The package includes a library of media containing all the necessary data taken from NIST databases, so the user only has to choose the media from this library. Adding new media to the library is also quite easy.
%%JR: detalles de geometría 
%Presently, only three object geometries (i.e., sphere, cylinder and orthohedron) are supported. Any object can be divided into voxels in Cartesian coordinates, although spherical and cylindrical voxelizations are also available for objects having these symmetries. The user only needs to select the geometry, input the dimensions of the object, and give the number of voxels along each dimension.
Presently, only three object geometries are supported: sphere, cylinder and orthohedron. In the two latter cases, the object is assumed to be oriented along the z axis. The object can be divided into two regions of different materials, with the boundary being a horizontal plane, a coaxial cylindrical surface (only for cylindrical objects), or a concentric spherical surface (only for spherical objects). Any object can be divided into voxels in Cartesian coordinates, although spherical and cylindrical voxelizations are also available for objects having these symmetries. The user only needs to select the geometry, input the dimensions of the object in the corresponding coordinates, give the number of voxels along each dimension of the object and, when applicable, set the position of the boundary.
%%

%%JR: detalles del beam. La última frase se lleva al siguiente párrafo.
%The beam of particles is assumed to enter the bottom surface of the object, which is always oriented along the z-axis. Both parallel and divergent beams can be used and a set of predefined energy spectra are available. For photon beams, the tracking of secondary electrons can be turned on and off. All these simplifications and default options provide a simple but flexible simulation framework.
The beam of particles is modeled in a simple way with a few parameters. The user may choose between a point source at any position inside the object and a parallel or divergent beam entering the bottom face of the object (only for a cylinder or an orthohedron). In the latter case, the user can set the angle of incidence, the position of the focal point (for a divergent beam) or the center of beam at the entrance plane (for a parallel beam), the shape of the beam (i.e., circular or rectangular) and its size at the entrance. Whether the beam particles are photons or electrons, a set of predefined energy spectra are available (e.g., monoenergetic, exponential, flat, Gaussian or given by a text file). For photon beams, the tracking of secondary electrons can be turned on and off.

All these simplifications and default options provide a simple but flexible simulation framework. In Fig. \ref{fig:legpy}, a sample of code that executes a simulation with LegPy and a couple of output plots are shown. The object dubbed ``result'' in this example stores the results of the simulation and has several methods to plot the energy deposit distribution, the particle tracks and histograms of various relevant parameters. The package includes additional functions to analyze the simulation results. All the results shown in this paper have been obtained with LegPy.

\section{Validation of the algorithm}
\label{sec:validation}
In order to check the validity of the code, a number of tests were made to compare our results with experimental data and those from other MC codes. In particular, we carried out systematic comparisons with the well-established code PENELOPE \cite{penelope}. These tests were done for the geometries currently supported by LegPy, i.e., cylinder, orthoedron, and sphere, and for the cases of a single medium and two media. Monoenergetic beams of photons and electrons in the range $0.1 - 5.0$~MeV were used.

%%FAM stastistical fluctuations
As will be shown below, for these comparisons the number of events simulated with both LegPy and PENELOPE were in general sufficient to reduce statistical fluctuations below the observed systematic deviations due to the simplifications of the LegPy algorithm. Number of events in between $5\times 10^5$ and $1\times 10^7$ were enough, depending of the case, to this end.
%%

%% FAM: tracking parameters
The energy threshold for transporting both photons and electrons has been set to 1.0 keV for both LegPy and PENELOPE. In addition, PENELOPE uses another tracking parameters for the transportation of electrons. For these comparisons, we have set the following values $C_1=C_2=0.05$ and $W_{\rm cc}=W_{\rm cr}=250$~eV, as suggested in the PENELOPE manual \cite{Salvat}.

\subsection{Photon beams}
\label{sec:photons}
For these tests, the LegPy simulations were carried out without the transportation of secondary electrons, that is, the electrons generated by a photoelectric absorption or a Compton effect are assumed to be absorbed depositing their kinetic energy in the production point. In the first set of tests, the dose in depth calculated with LegPy for several media at various energies was compared with the results from PENELOPE and MCNP5. Other tests were focused on the angular distribution of escaping photons and the energy spectra of absorbed energy relevant to the characterization of scintillators.  

The dose in depth comparisons were done through the buildup factor $B$ for a point source in an infinite medium, which is defined as
\begin{equation} \label{eq:B}
B = \frac{D_{\rm r}}{D_{\rm t}}
\end{equation}
where $D_{\rm r}(R)$ is the ``real'' dose as a function of the distance from the source $R$, which can be calculated by simulation. The ``theoretical'' dose $D_t(R)$, that is, the expected one if secondary radiation (e.g., Compton scattered photons) is ignored, can be calculated from the expression
\begin{equation} \label{eq:Dt}
D_{\rm t}(R) = \frac{1}{4\pi R^2}  \exp\left({-\mu R}\right) \left( \frac{\mu_{\rm en}}{\rho} \right) E
\end{equation}
where $\mu$ and $\mu_{\rm en}/\rho$ are respectively the linear attenuation coefficient and the mass absorption coefficient of the medium at the photon energy $E$, tabulated in \cite{nist_ph}.

Buildup factors are very relevant in engineering materials for shielding calculations. A database (ANSI/ANS-6.1.1, 1991) is available in \cite{ansi-ans91} and their values are confirmed by MC simulations with EGS4 \cite{hirayama}, \cite{UEI}. These parameters are also of the most relevance in medical physics \cite{Manohara}, \cite{kadri}. 

\begin{figure}[h]
\centering
\includegraphics[width=\linewidth]{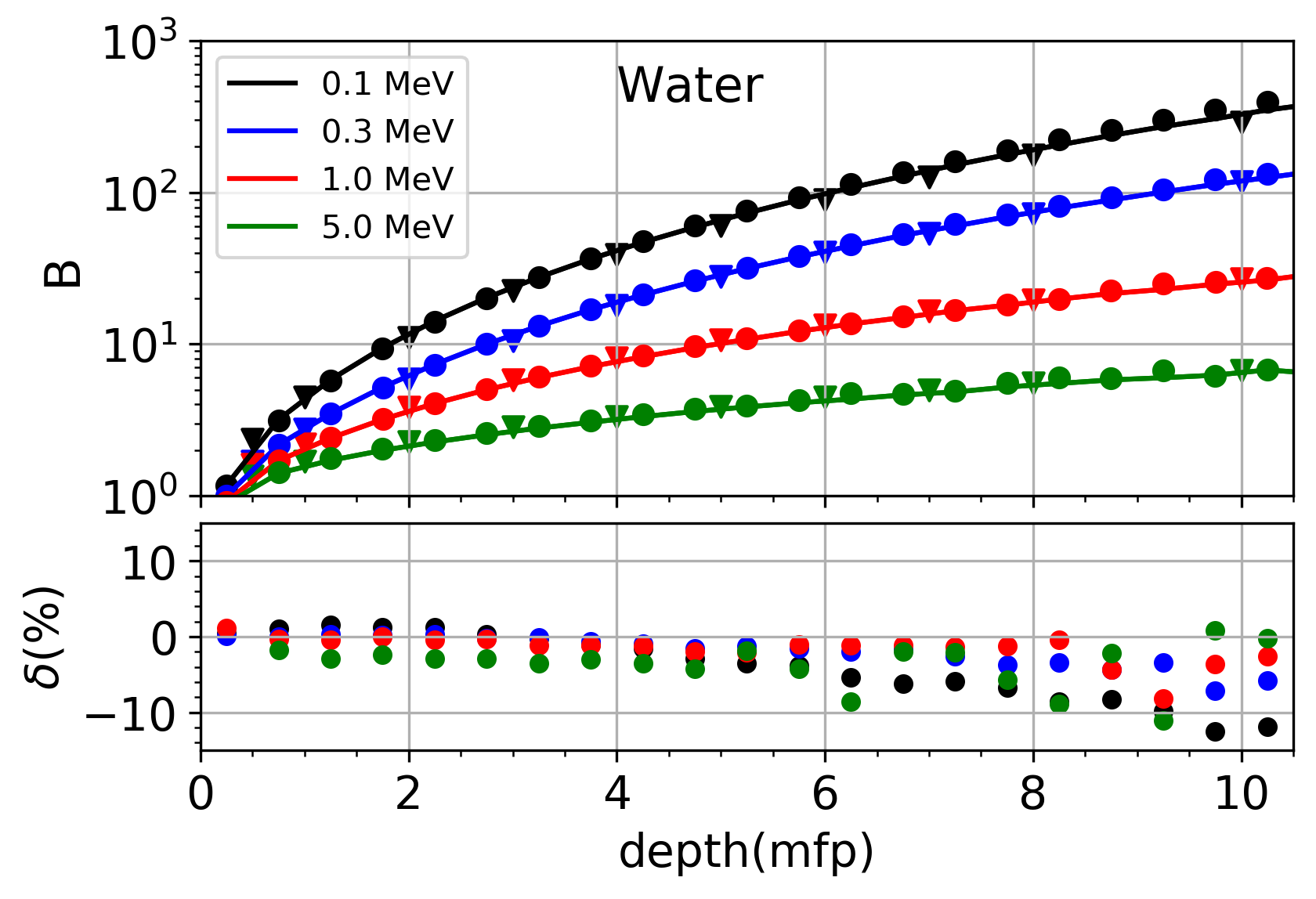}
\caption{In the upper plot the buildup factors versus depth (in mean free path units) in water obtained with LegPy (continuous lines) are compared with those obtained with PENELOPE (full circles) and MCNP5 (triangles). The deviations of LegPy results with respect to those from PENELOPE are shown in the lower plot.}
%%FAM he actualizado las 3 figuras. A los png les numero como estan las figuras en el paper. 
\label{fig:B_w}
\end{figure}

\begin{figure}[h]
\centering
\includegraphics[width=\linewidth]{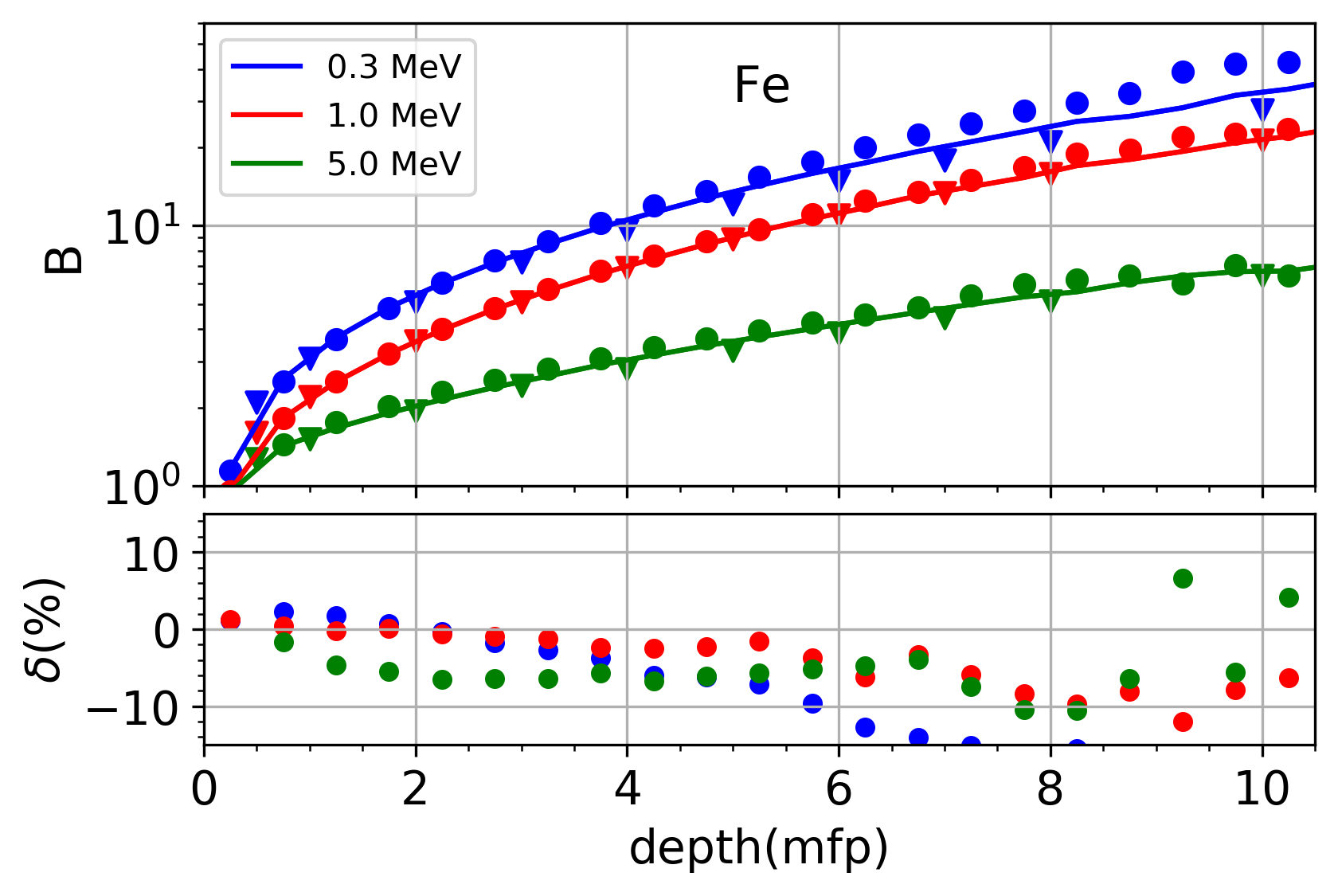}
\caption{Same as figure \ref{fig:B_w} for iron.}
\label{fig:B_Fe}
\end{figure}

\begin{figure}[h]
\centering
\includegraphics[width=\linewidth]{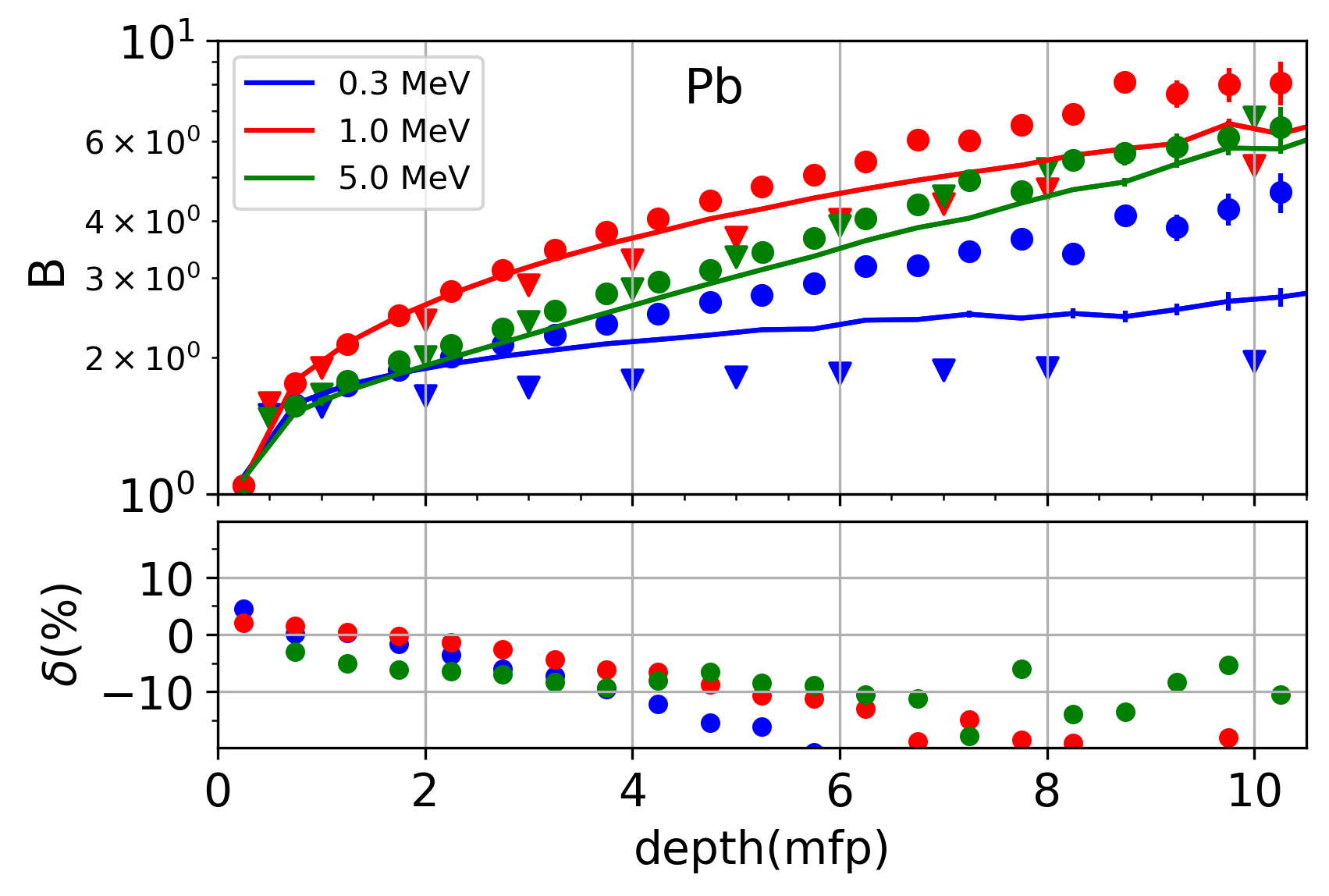}
\caption{Same as figure \ref{fig:B_w} for lead.
%%FAM statistics
The small error bar at large depth accounts for the statistical fluctuations of the simulations}
\label{fig:B_Pb}
\end{figure}

%%JR: Medio infinito
%Using LegPy, we calculated $B(R)$ for water, iron, and lead at energies of 0.1, 0.3, 1.0, and 5.0 MeV. We repeated the simulations using PENELOPE and recalculated the buildup factors.
We calculated $B(R)$ for water, iron, and lead at energies of 0.1, 0.3, 1.0, and 5.0 MeV using both LegPy and PENELOPE. To do that, we simulated a point source at the center of a sphere of large radius, i.e., $15$~mfp of the corresponding material.
In addition, we used the updated ANSI/ANS-6.1.1 buildup factor database \cite{ansi_up_low}, \cite{ansi_up_high} obtained by MCNP5 simulations \cite{mcnp5}. The results are compared in figures \ref{fig:B_w} - \ref{fig:B_Pb}. Note that comparing the buildup factor is equivalent to comparing the dose, assuming that the same $\mu$ and $\mu_{\rm en}/\rho$ coefficients are used.

%%FAM Statitical fluctuations
Statistical fluctuations for depths smaller than 5 mfp are well below 1\% for all cases shown here and for both simulations, LegPy and PENELOPE. The fluctuations grow at larger depths, although the error bars can only be appreciated in a few points and are significantly smaller than the systematic deviations of LegPy due to our simplifications. 


As can be seen in Fig.~\ref{fig:B_w} (upper plot), the buildup factors for water obtained with LegPy are in good agreement with those obtained with both PENELOPE and MCNP5. The deviations of LegPy results with respect to those from PENELOPE (lower plot) are at the level of less than 5\% for $R < 4$~mfp, growing up to about 
%%FAM
%15\%
$10\%$
at 10~mfp. Notably, the agreement is very good even at 5.0~MeV, confirming that the impact of ignoring pair creation and Bremsstrahlung on the calculation of the dose in depth is very weak in light elements even at energies of a few MeV.

Results for iron at 0.3, 1.0 and 5.0 MeV are shown in Fig.~\ref{fig:B_Fe}. Again, the simplifications made in LegPy do not seem to have a relevant impact in the calculation of the dose at these energies. At 0.1 MeV (not shown in the figure), LegPy also reproduces the result from PENELOPE up to about 4~mfp but underestimates it by 30\% at 10~mfp. This deviation is attributed to the simplified treatment of the coherent scattering in LegPy.

The results for lead at 0.3, 1.0, and 5.0 MeV are shown in Fig.~\ref{fig:B_Pb}. LegPy is in agreement
%%JR: definir agreement
($\delta<10\%$)
with PENELOPE up to about 4~mfp with larger deviations (up to 30\%) at large depths. Again, the effect of ignoring pair creation and Bremsstrahlung is not very relevant in the depth dose calculation even at 5 MeV. However, we found very large deviations with respect to both PENELOPE and MCNP5 at 0.1 MeV, which are attributed to ignoring X-ray fluorescence. It is well known~\cite{subbaiah} that, for heavy atoms, X-ray fluorescence leads to a dramatic increase in buildup factors for photon energies slightly above the K edge of the photoelectric cross section. Indeed, the value of $B(R)$ obtained by PENELOPE is larger than that obtained with LegPy by a factor of 6 (50) at a depth of 5~mfp (10~mfp).

\begin{figure}[h]
\centering
\includegraphics[width=\linewidth]{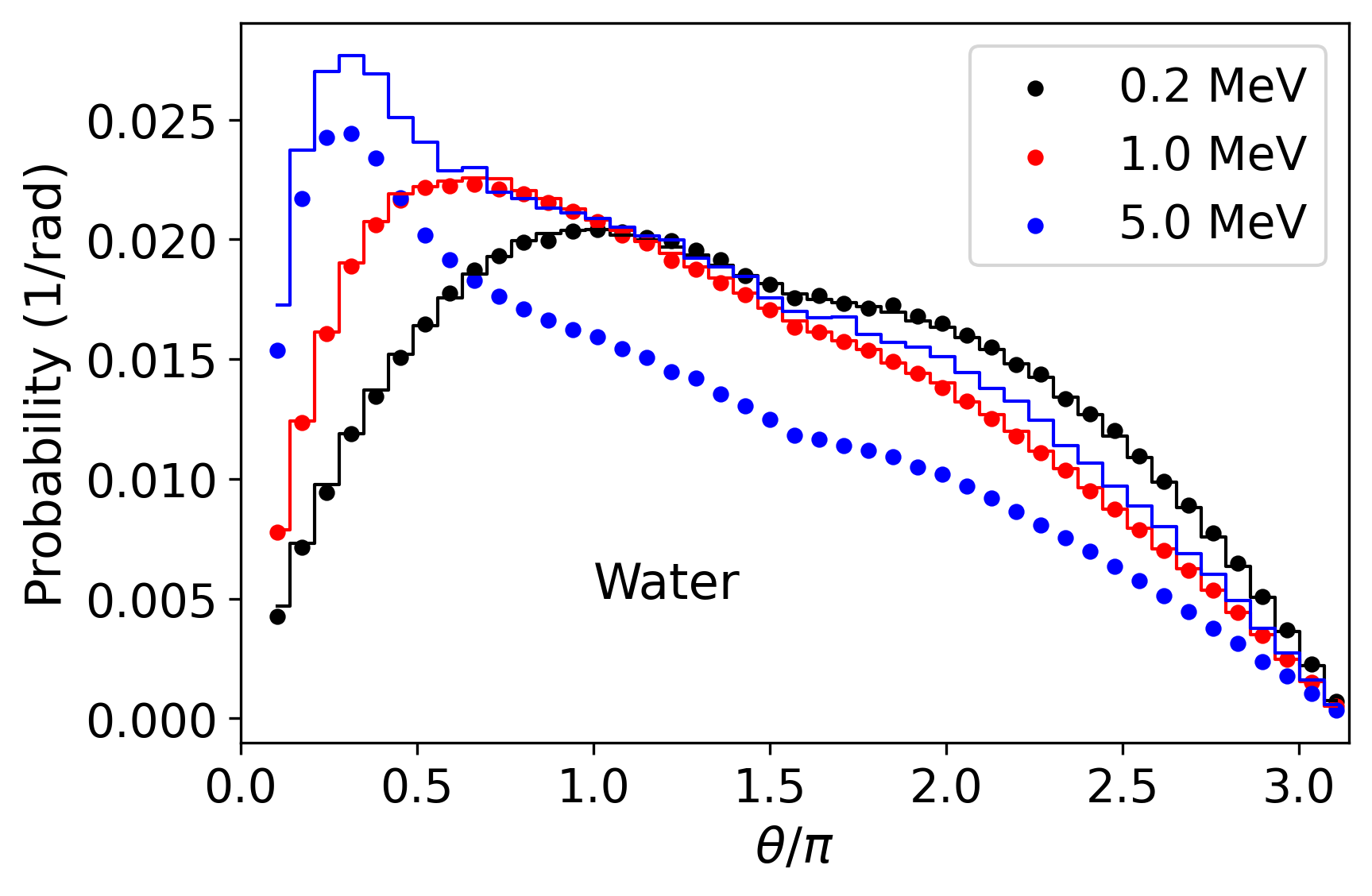}
\caption{Angular distribution of outgoing photons for 0.2, 1.0 and 5.0 MeV pencil beams on a water cylinder of one mean free path size. Results from LegPy (points) are compared with those from PENELOPE (continuous line). See text for details.}
\label{fig:ang_w}
\end{figure}
\begin{figure}[h]
\centering
\includegraphics[width=\linewidth]{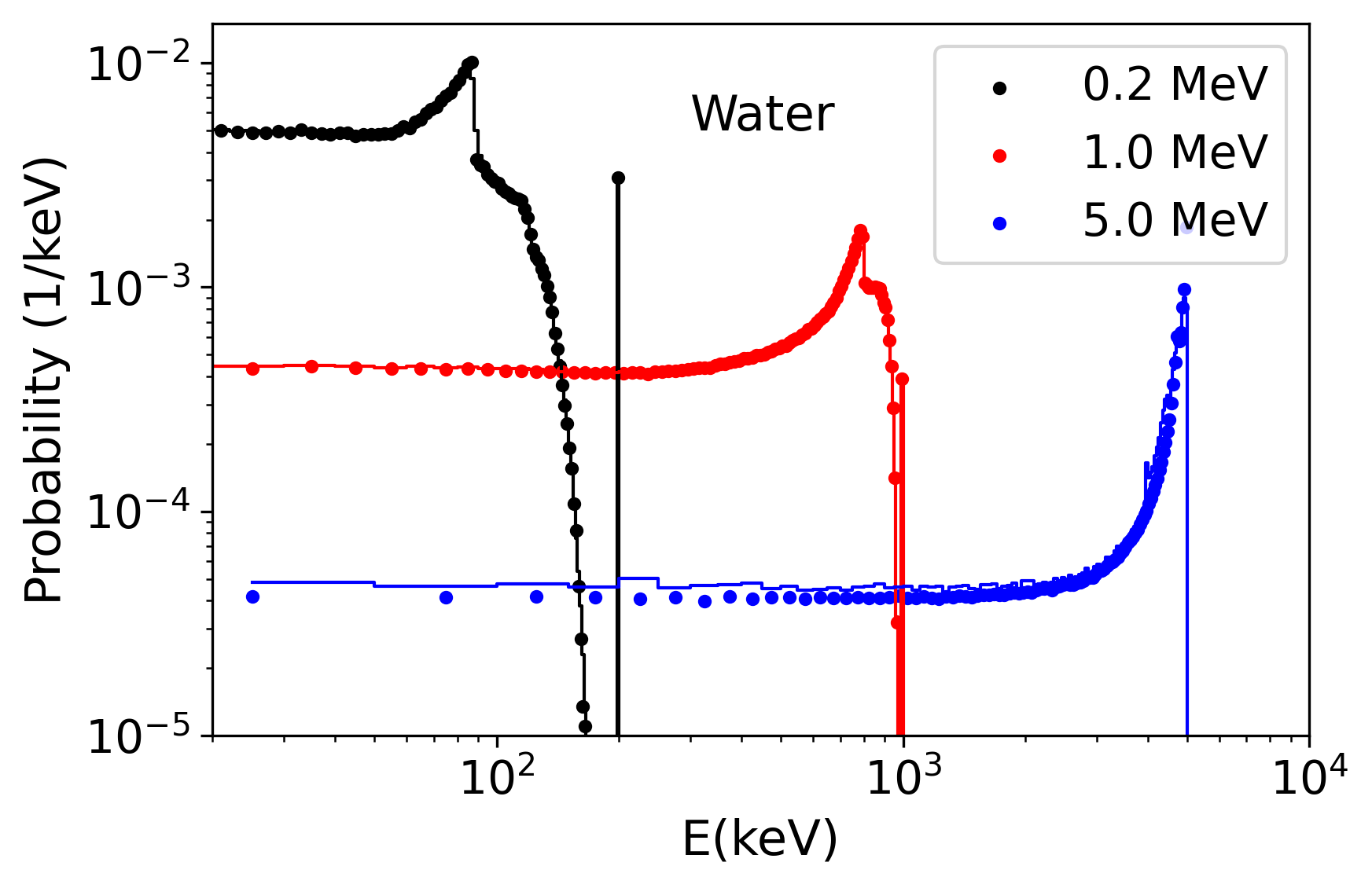}
\caption{Spectra of absorbed energy for the same simulation tests as used in Fig. \ref{fig:ang_w}.}
\label{fig:eab_w}
\end{figure}
\begin{figure}[h]
\centering
\includegraphics[width=\linewidth]{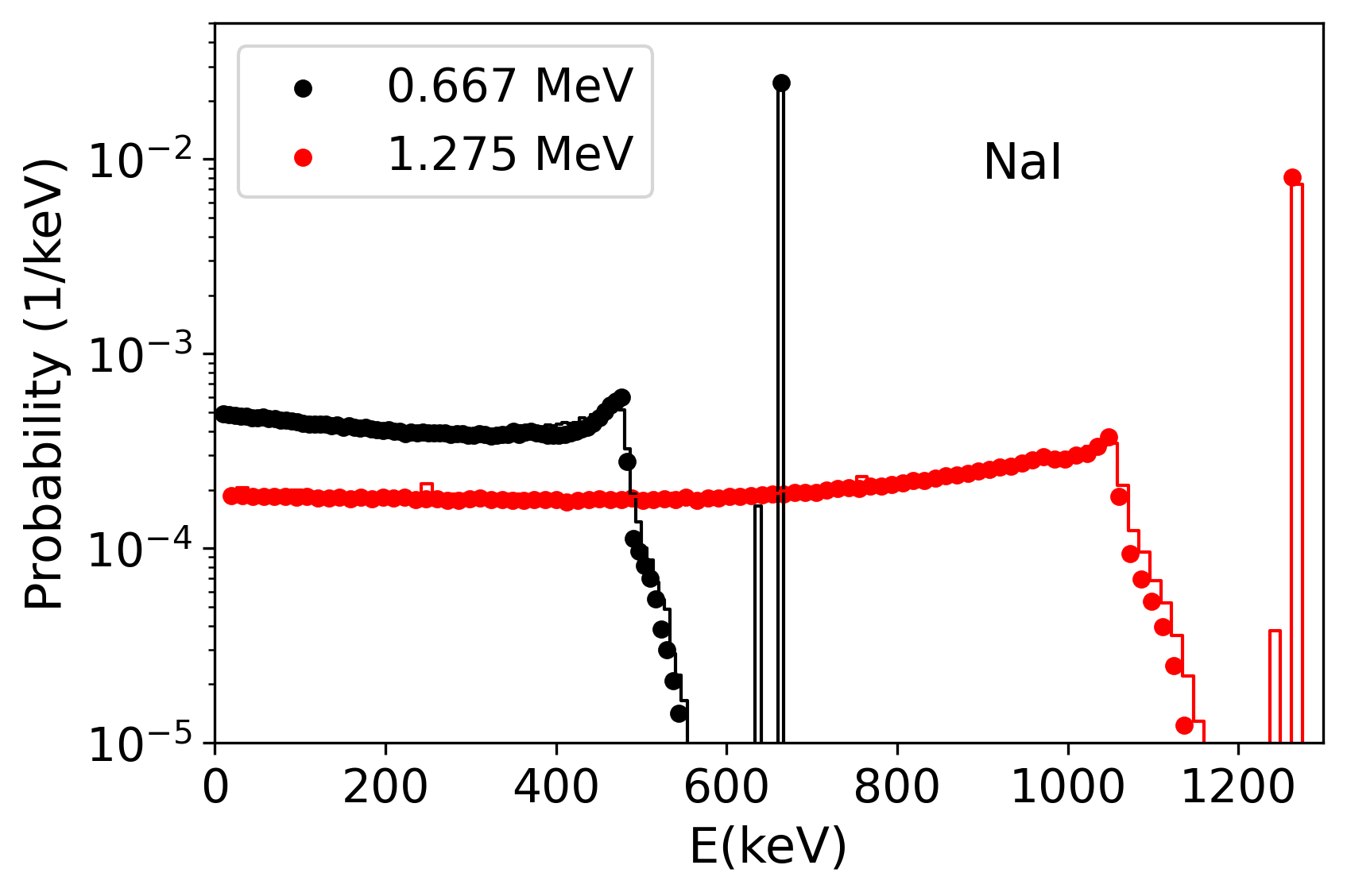}
\caption{Spectrum of absorbed energy for isotropic sources of 0.667 and 1.275 MeV on a NaI cylinder. Results from LegPy (points) are compared with those from PENELOPE (continuous line). See text for details.}
\label{fig:eab_nai_667}
\end{figure}

We carried out another set of tests for the angular distribution of photons escaping the medium. We simulated pencil beams of various energies traversing a cylinder of water with both height and diameter equals to one mfp at the corresponding energy. The LegPy results for 0.2, 1.0, and 5.0~MeV are compared with those from PENELOPE in Fig.~\ref{fig:ang_w}. 
%%FAM statistical fluct and agreement
Statistical fluctuations in both simulations are smaller than or about $1\%$, except for angles close to 0 or 180 degrees because of the small solid angle. We found that the agreement is very good up to 2.0~MeV, with deviations within the above statistical fluctuations, while LegPy results deviate significantly from those from PENELOPE at higher energies (up to about $30\%$ at 5~MeV). From these simulations, we also calculated the distribution of energy absorbed by the medium. The results are shown in Fig.~\ref{fig:eab_w}. The agreement is very good in the photopeak and in the whole Compton profile before the sharp decrease at the edge, with deviations smaller than $2\%$ for energies up to 2~MeV. These deviations grow up to about $15\%$ at 5~MeV. In the deep Compton edge deviations are hidden by large statistical fluctuations. On the other hand,
LegPy does not reproduce some spectral features associated with processes ignored in our algorithm, e.g., escape of annihilating photons subsequent to pair creation.

An interesting test was done for isotropic beams of 0.667 and 1.275 MeV (emulating radioactive point sources of $^{137}$Cs and $^{22}$Na) at a distance of 2 cm from a NaI cylinder with both diameter and height of one mean free path (4.34 cm for $^{137}$Cs and 5.32 cm for $^{22}$Na). As can be seen in Fig. \ref{fig:eab_nai_667}, 
%%JR: definir agreement y reescribir un poco la frase
the main features of the spectrum of absorbed energy obtained with LegPy, i.e., the shape of the Compton profile and the height of the photopeak, are in good agreement with the one obtained with PENELOPE (i.e., deviations $\lesssim 5\%$).
However, PENELOPE predicts a small peak shifted by about 30 keV from the photopeak due to the escape of X-rays produced after the photoelectric absorption in the K shell of Iodine. This feature, which is generally irrelevant for the characterization of a scintillator,  cannot be reproduced by LegPy because X-ray fluorescence is ignored in our algorithm.

\subsection{Electron beams}
\label{sec:electrons}
For the validation of LegPy for the transportation of electrons, the dose in depth, the backscattering coefficient and the angular distribution of backscattered electrons were obtained for
%%JR: tracking parameters
%a pencil electron beam 
a pencil beam of monoenergetic electrons
going along the axis of a semi-infinite cylinder (i.e.,  larger than the corresponding CSDA range). Different media were tested in the energy range $0.1 - 5.0$~MeV.
%%JR: tracking parameters
The step length for the transportation of electrons was set to $1\%$ of the CSDA range.
As in the previous section, we repeated the simulations with PENELOPE to compare with LegPy. The simulation results were also compared with experimental data on energy deposition of electron beams on different media and backscattering coefficients, which are available in the literature for a wide energy range \cite{sandia_el_1}, \cite{sandia_el_2} and have been used since long ago to benchmark PENELOPE \cite{Sempau} and  GEANT4 \cite{Kadri}, \cite{P_Arce}. 

\begin{figure}[h]
\centering
\includegraphics[width=\linewidth]{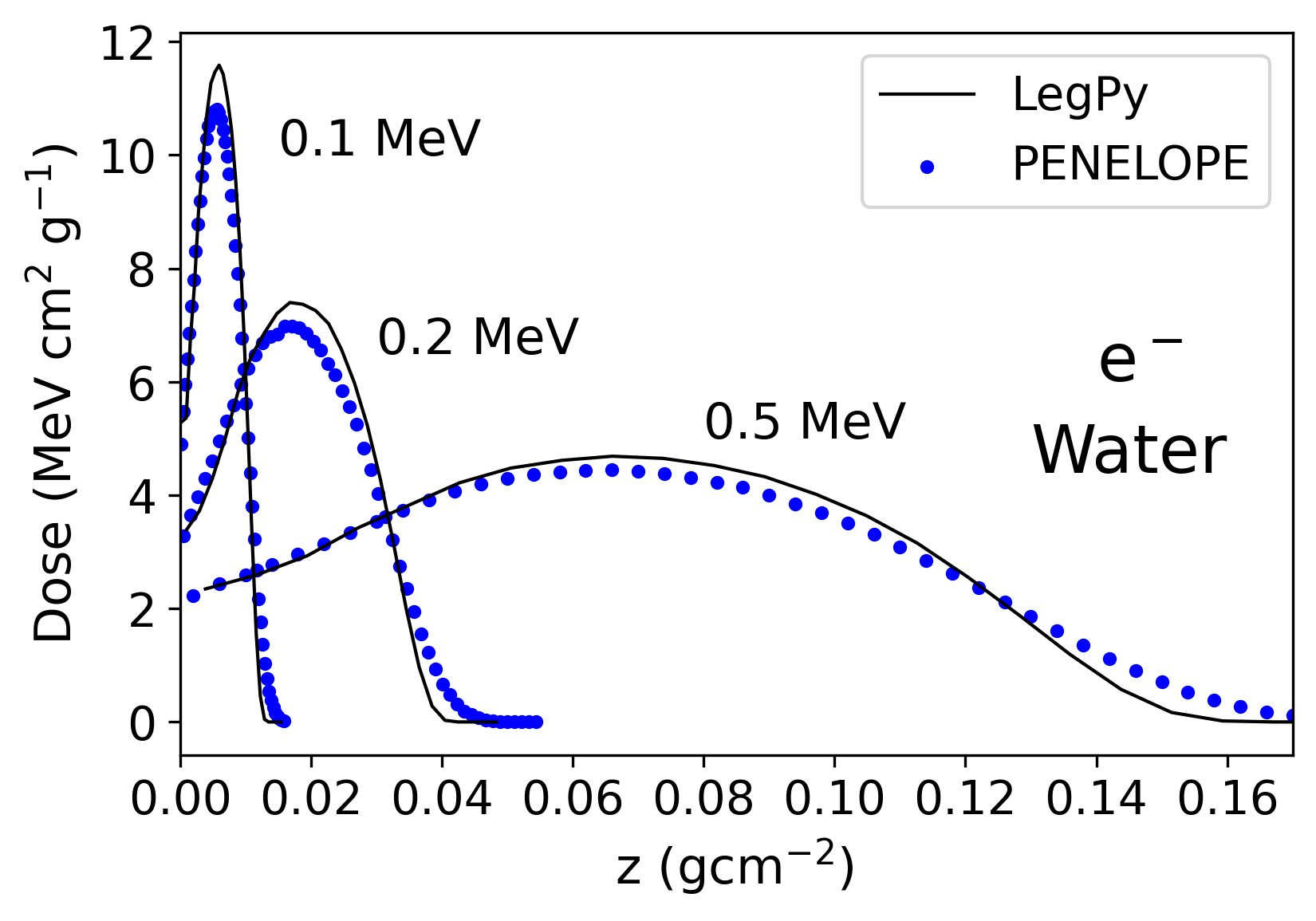}
\caption{Depth dose for electron beams of 0.1, 0.2, and 0.5 MeV on water. The results from LegPy (continuous lines) are compared with those from PENELOPE (full blue circles).}
\label{fig:w_0.1_0.5}
\end{figure}

\begin{figure}[h]
\centering
\includegraphics[width=\linewidth]{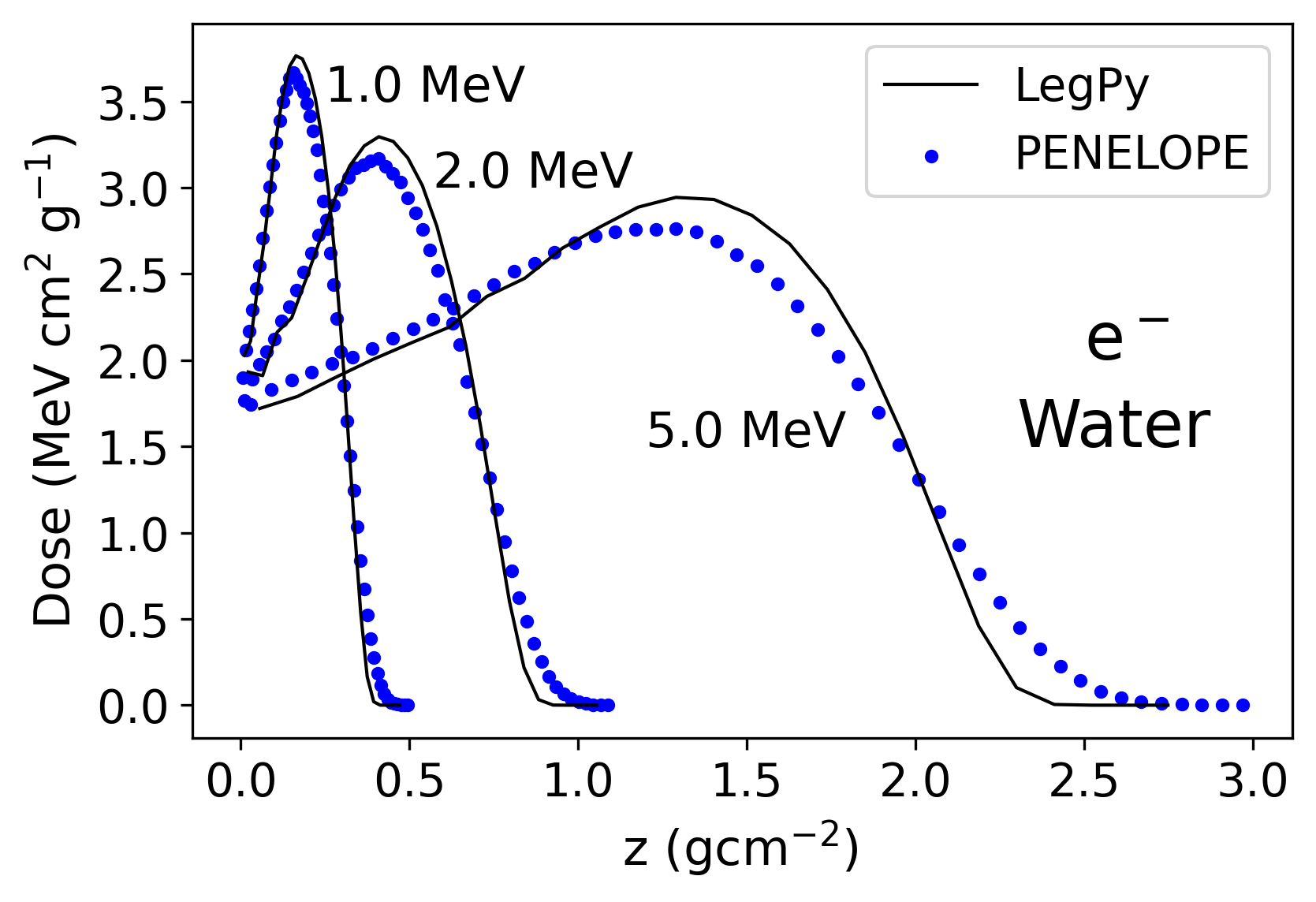}
\caption{Same as Fig. \ref{fig:w_1_5} for 1.0, 2.0 and 5.0 MeV.}
\label{fig:w_1_5}
\end{figure}

In figures \ref{fig:w_0.1_0.5} and \ref{fig:w_1_5}, depth dose in water for electrons at several energies in the range $0.1 - 5.0$ MeV obtained with LegPy are compared with the results from PENELOPE. As explained above, the fluctuations in energy deposition are ignored in our simplified algorithm in such a way that the path length of all electrons equals the CSDA range. As a consequence, the dose at depth beyond the CSDA range obtained with LegPy is zero while the energy is spread to larger distances for the PENELOPE results. Apart from this discrepancy, LegPy reproduces reasonably the results from PENELOPE with deviations in the maximum of the dose in depth function smaller or around 10\%. These discrepancies are a consequence of the approximations made in the multiple scattering and collisional energy losses in our algorithm.

\begin{figure}[h]
\centering
\includegraphics[width=\linewidth]{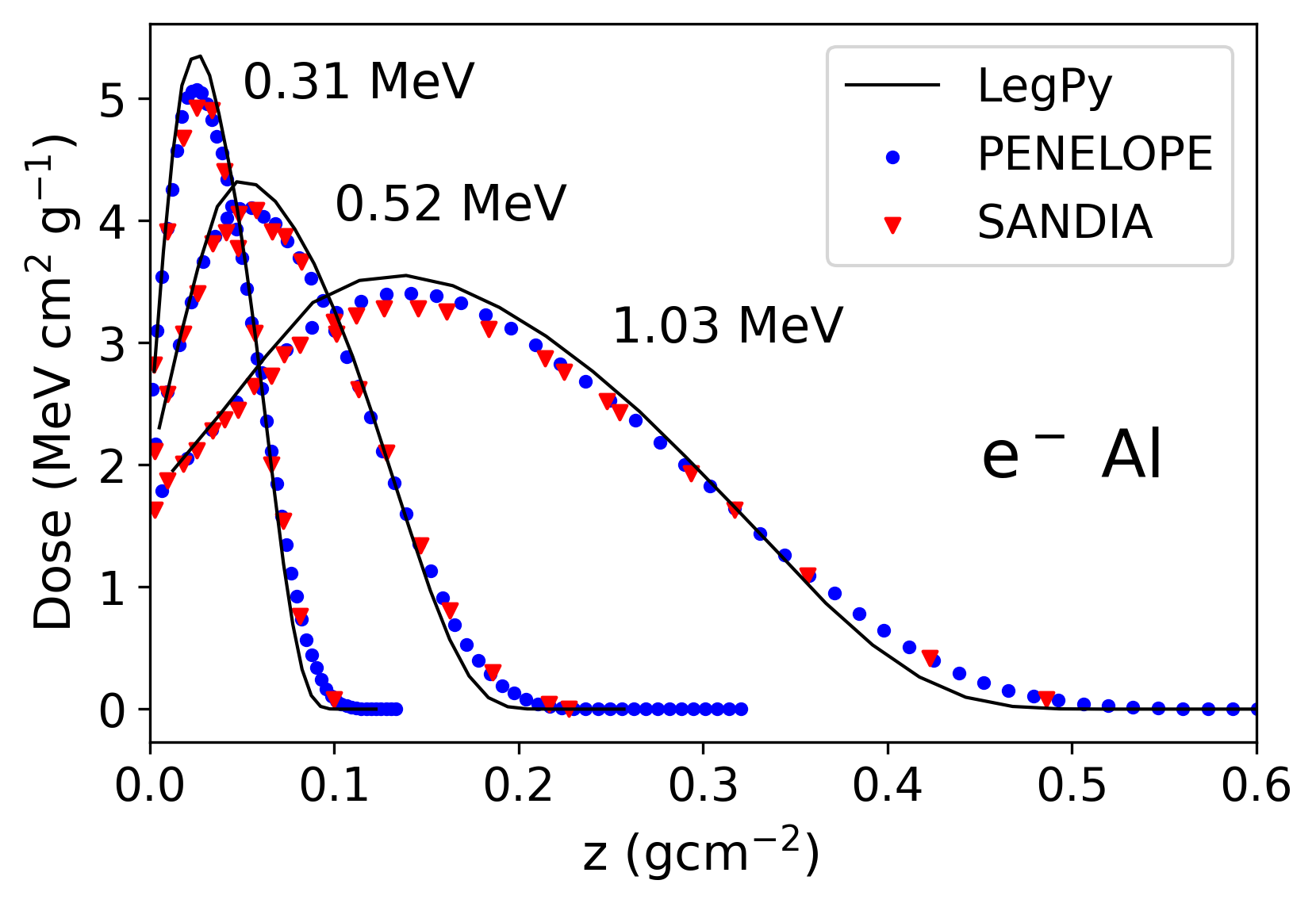}
\caption{Depth dose for electron beams of 0.31, 0.52, and 1.03 MeV on aluminum. The results from LegPy (continuous lines) are compared with those from PENELOPE (full blue circles) and experimental data from \cite{sandia_el_1} (red triangles).}
\label{fig:Al_0.3-0.5-1.0}
\end{figure}
\begin{figure}[h]
\centering
\includegraphics[width=\linewidth]{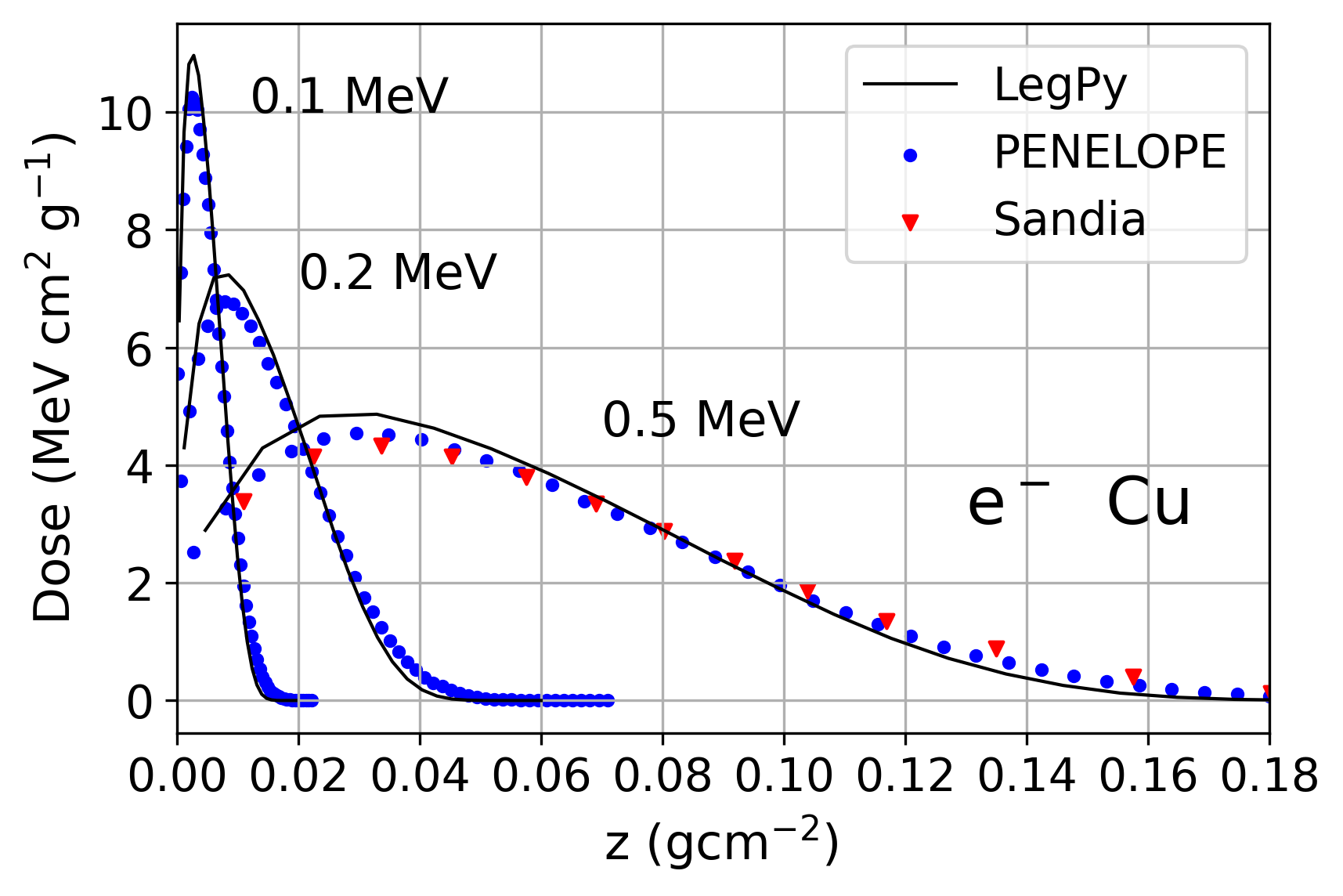}
\caption{Same as Fig. \ref{fig:Al_0.3-0.5-1.0} for 0.1, 0.2, and 0.5~MeV electrons on copper.}
\label{fig:Cu_0.1_0.5}
\end{figure}
\begin{figure}[h]
\centering
\includegraphics[width=\linewidth]{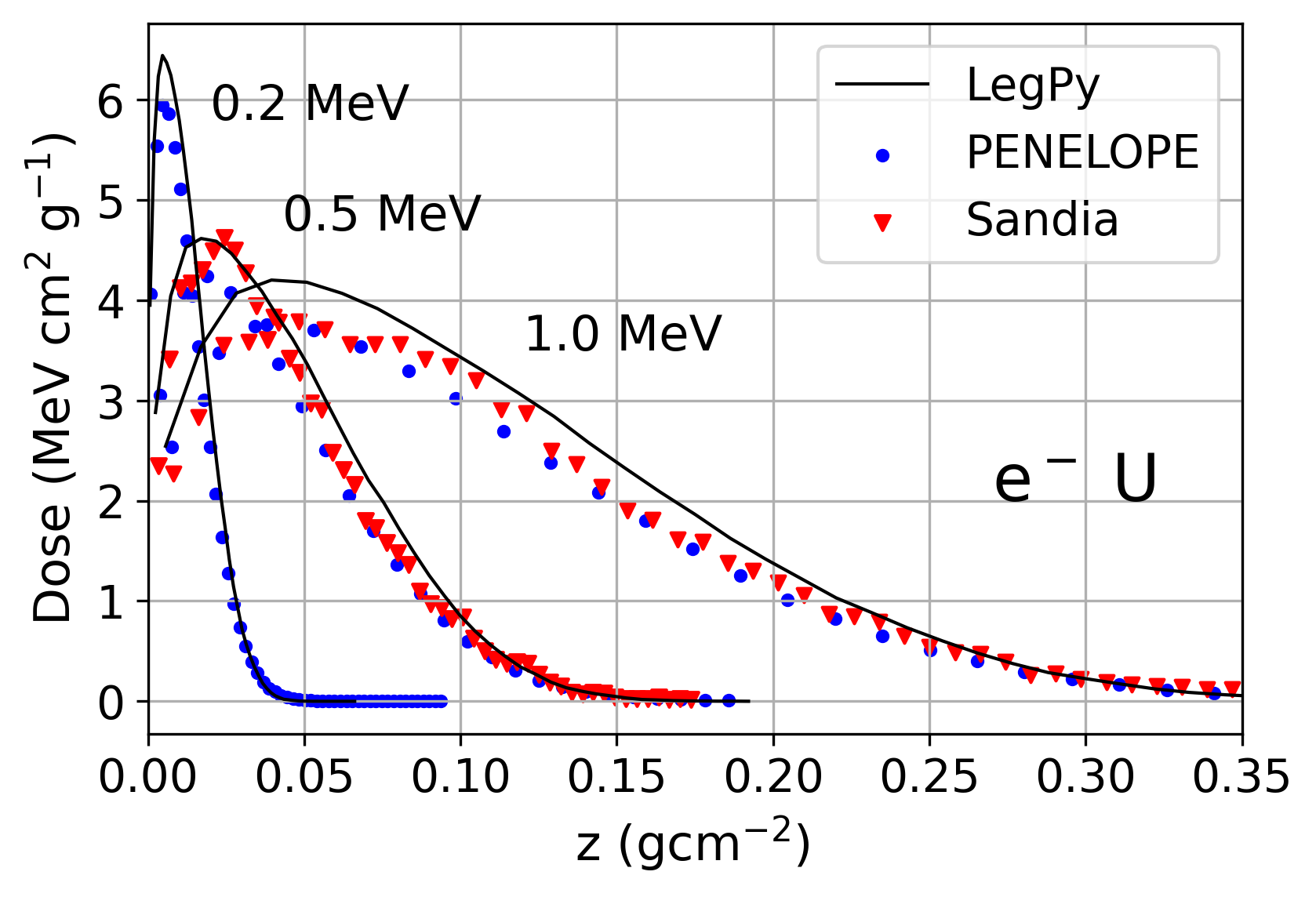}
\caption{Same as Fig. \ref{fig:Al_0.3-0.5-1.0} for 0.2, 0.5 and 1.0~MeV electrons on uranium.}
\label{fig:U_0.2_1.0}
\end{figure}

The results for aluminum are shown in Fig.~\ref{fig:Al_0.3-0.5-1.0}. LegPy reproduces the depth dose of PENELOPE with
%%FAM
%an accuracy 
similar deviations to those
obtained for water. Available experimental data at 0.31, 0.52, and 1.03~MeV \cite{sandia_el_1} are in very good agreement with PENELOPE.
%%FAM No hace falta esto
%confirming the reasonable accuracy of LegPy.
Similar results were found up in the whole range between 0.1 and 5.0~MeV.

The results for copper are shown in Fig.~\ref{fig:Cu_0.1_0.5}. LegPy also reproduces 
%%FAM
%well 
reasonably
the dose in depth at energies below 1~MeV. The discrepancy turns out significant at 5.0~MeV ($16\%$ at the depth dose maximum), very likely due to ignoring the Bremsstrahlung production. For very heavy elements, such as uranium, this discrepancy exists even at lower energies, as expected (Fig.~\ref{fig:U_0.2_1.0}).

\begin{figure}[h]
\centering
\includegraphics[width=\linewidth]{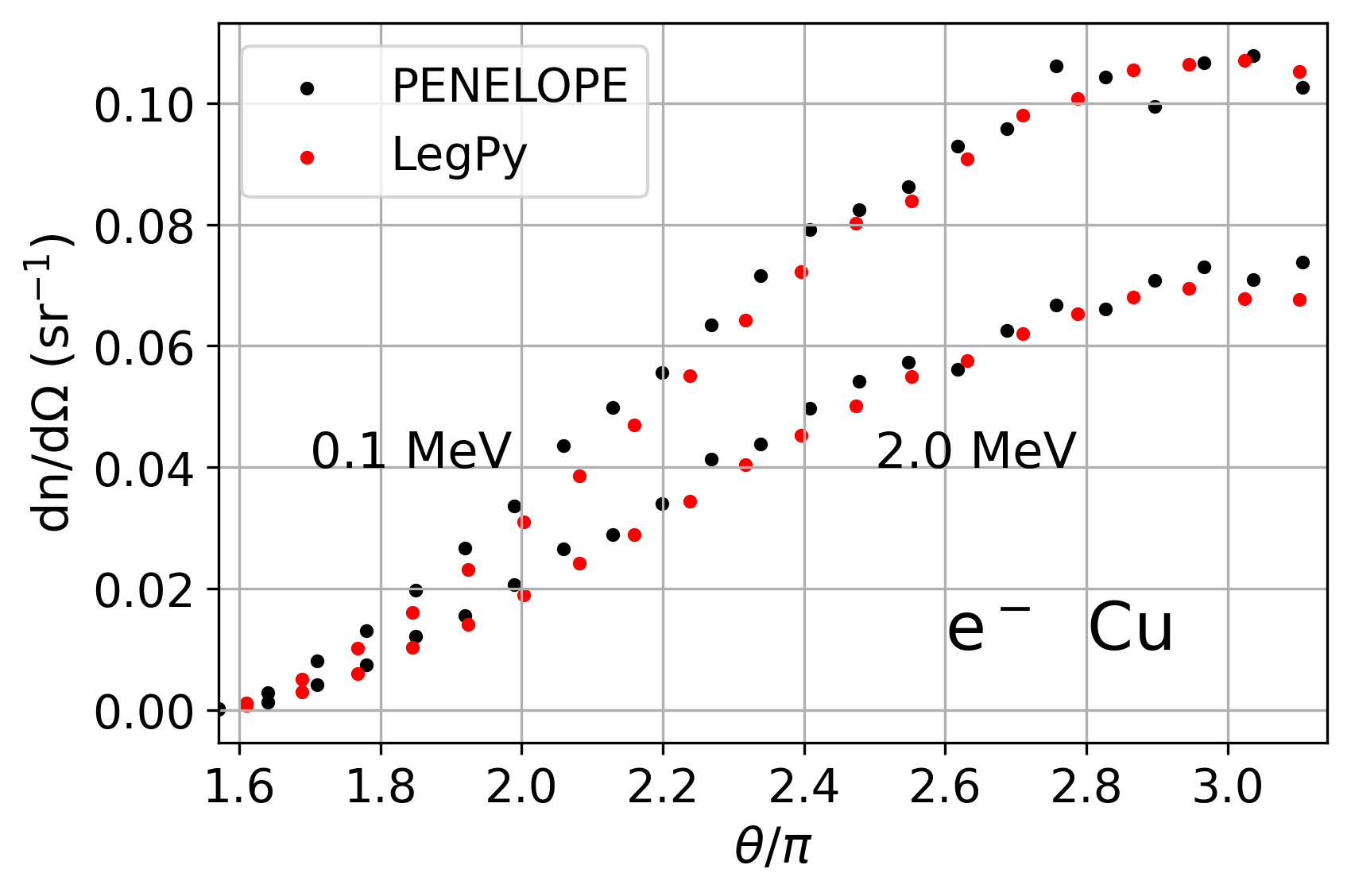}
\caption{Angular distribution of backscattered electrons in copper. The results from LegPy (red points) are compared with those from PENELOPE (black points)}
\label{fig:Cu_bck}
\end{figure}

Electron backscattering in LegPy was also compared with the results from PENELOPE. In general, we found a good agreement in the shape of the angular distribution of backscattered electrons obtained with both simulation codes. As an example, we show the results for copper at 0.1 and 2.0~MeV in Fig.~\ref{fig:Cu_bck},
%%JR: definir agreement
where deviations are in general smaller than $5\%$.
On the other hand, LegPy tends to underestimate the backscattering coefficient $\eta$ (i.e., the ratio between incoming and backscattered electrons) for light elements. For instance, in water at 1.0~MeV, the PENELOPE result is $\eta = 3.0\%$ while it is half that value for LegPy. The agreement in the backscattering coefficient improves for heavier elements. For aluminum at 1.03~MeV, LegPy gives $\eta =8.0\%$ to be compared with the value of $\eta = 9.5\%$ obtained with PENELOPE, which is in good agreement with the experimental value of $9.2\%$ \cite{sandia_el_2}. For copper at 2.0~MeV (Fig.~\ref{fig:Cu_bck}), LegPy gives  $\eta = 28\% $ while PENELOPE gives $\eta = 29.5\% $. For uranium at 1 MeV, the PENELOPE result is $\eta = 51\%$ versus $\eta = 44\%$ from LegPy.

\subsection{Beams in a two media object}
\label{sec:2M_ph}
\begin{figure}[h]
\centering
\includegraphics[width=\linewidth]{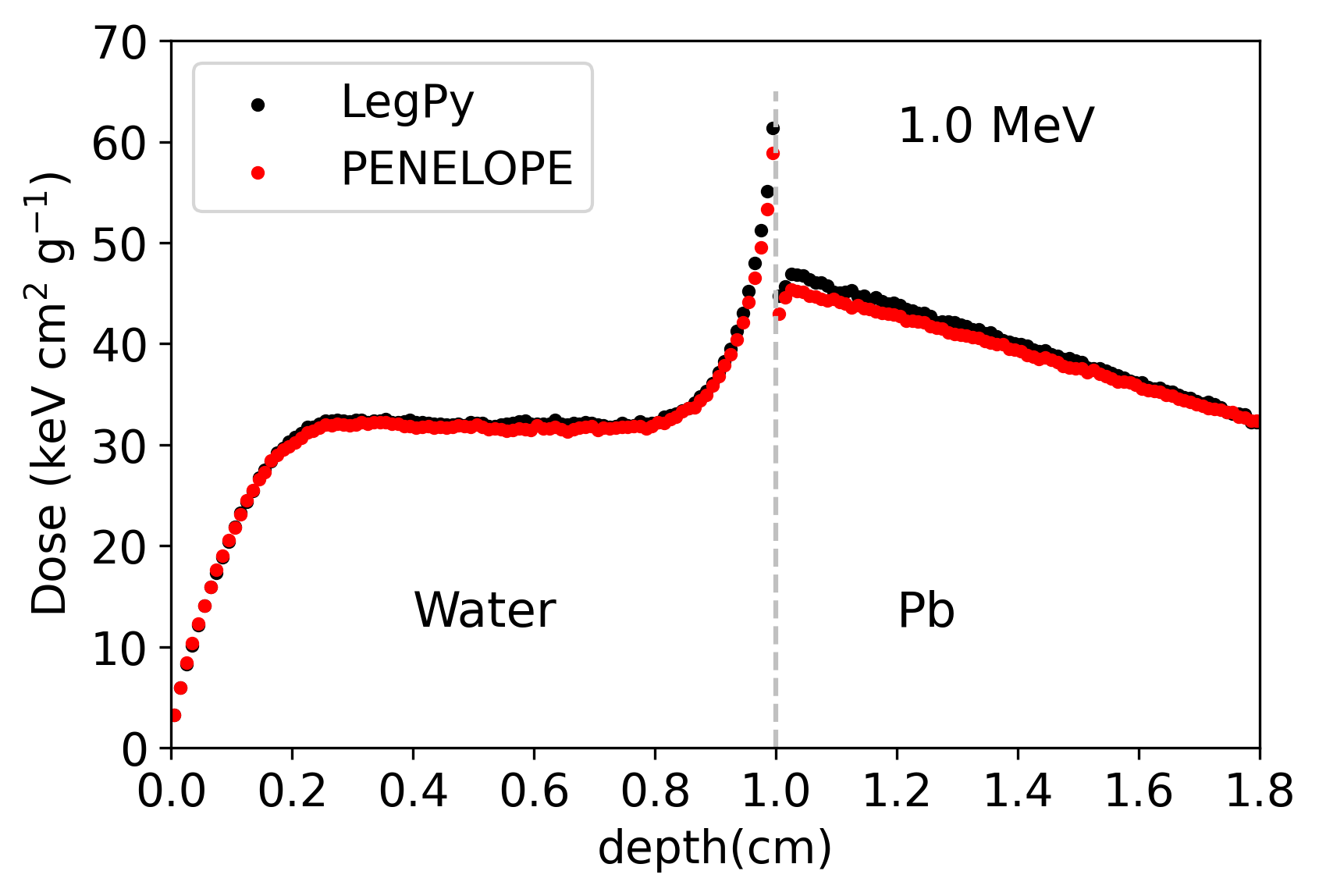}
\caption{Dose in depth for a beam of 1.0 MeV photons on a water - lead cylinder along the z axis. The dashed line indicates the position of the boundary between the two media. See text for details. The results from LegPy (black points) are compared with those from PENELOPE (red points)}
\label{fig:w-Pb_1.0}
\end{figure}

Several tests were made to check the performance of LegPy for transporting photons and electrons through an object composed by two different media. As an example, in Fig.~\ref{fig:w-Pb_1.0}, we show the dose in depth for a photon beam 
%%FAM esto estaba mal en la version anterior.
%of 1.0 MeV
of 10~cm diameter and 1.0~MeV
energy crossing along its axis a cylinder of 2.0 cm length
%%FAM
%with a 10.0 cm diameter
with a 100~cm diameter
(i.e., laterally infinite) to assure that all scattered radiation except the backscattered one is absorbed. The cylinder is made of water and lead, the boundary between the two media being at a depth of 1.0~cm.
%%JR: tracking parameters
The transportation of secondary electrons is included in the simulation with a step length of 7~$\mu$m and a voxel size of 90~$\mu$m in order to study the details at the boundary.
As can be seen in the figure, the dose in depth obtained with LegPy is in good agreement
%%FAM
with the result from PENELOPE. The average deviation is of about $1.5\%$ with a maximum value of $3\%$ around the interface surface. 
Note the relevant role of secondary electrons in the dose in depth. Electronic equilibrium is reached at about 2.5~mm and the backscattering in lead is very strong. We checked that similar agreements with PENELOPE are achieved at 0.3~and 5.0 MeV.

\begin{figure}[h]
\centering
\includegraphics[width=\linewidth]{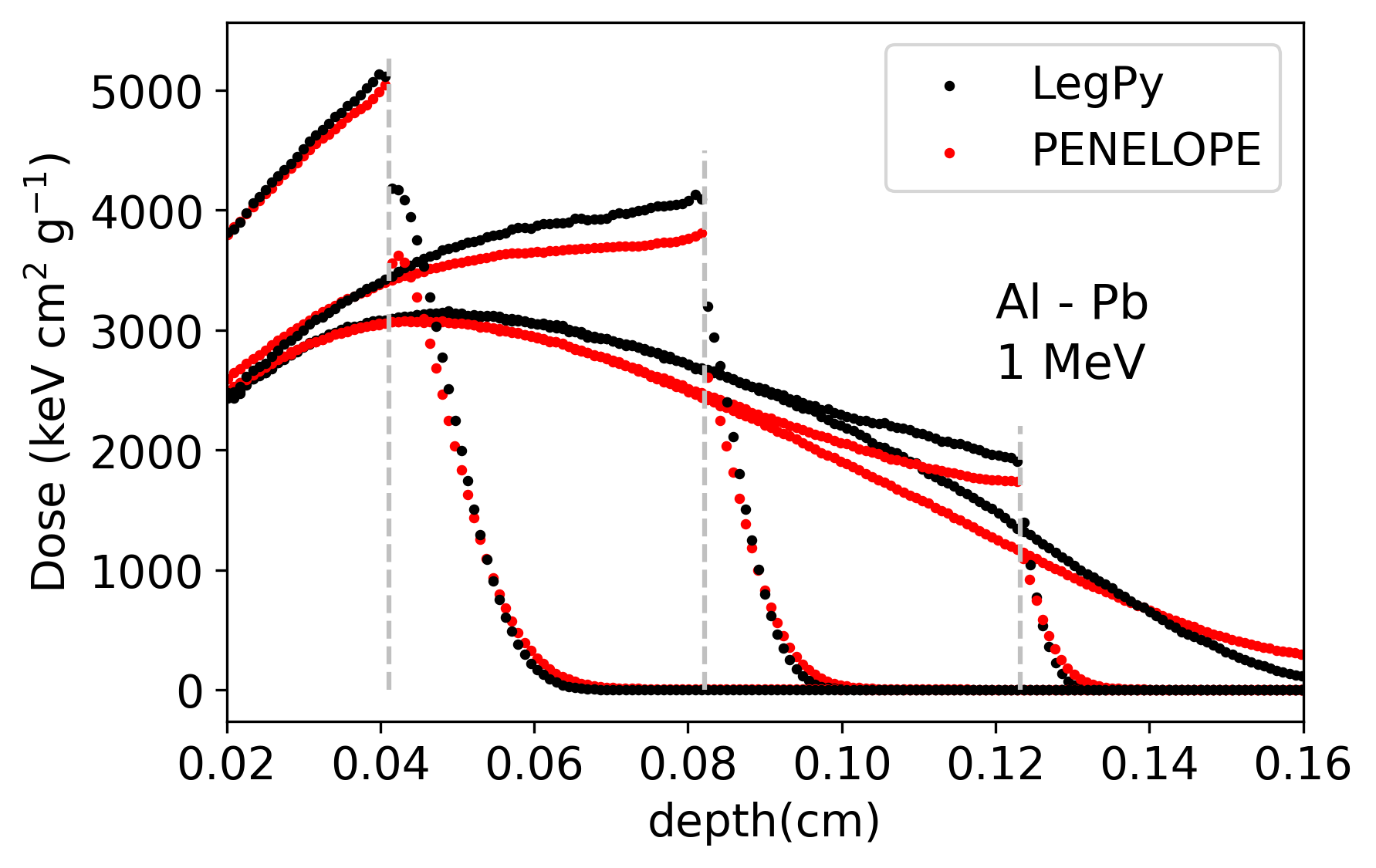}
\caption{Dose in depth for an electron beam of 1.0 MeV on an aluminum - lead cylinder along the z axis. The dashed lines indicate the position of the boundary between the two media. See text for details. The results from LegPy (red points) are compared with those from PENELOPE (red points)}
\label{fig:Al-Pb_1MeV}
\end{figure}

Tests for an electron beam traversing a two-media cylinder were also carried out. As an example, in Fig.~\ref{fig:Al-Pb_1MeV} it is shown the dose in depth for a pencil beam of 1.0~MeV electrons crossing a cylinder of 0.164~cm length and 0.20~cm diameter, made of aluminum and lead.
%%JR: tracking parameters
The step length was set to $1\%$ of the CSDA range of the electrons in lead.
Several cases were studied varying the depth $z_{\rm b}$ of the boundary between the two media. The results of the figure
%% JR: corregido
correspond
to $z_{\rm b}$ values of 0.0411, 0.082, and 0.123 cm as well as to the only-Al case. 
%%FAM
LegPy reproduces the features of the behavior of the dose around the interface due to backscattering. However deviations with respect to PENELOPE are up to $15\%$,
basically due to the approximations in the electron transportation, as already discussed in \ref{sec:electrons}.

\section{Conclusions}
\label{sec:conclusions}
In this paper, we have presented a simplified algorithm for the simulation of the passage of low-energy electrons and gamma rays through any medium. The algorithm has been realized in the Python package called LegPy available under an open source license.

The algorithm has been validated by comparing a set of results with the code PENELOPE as well as with available experimental data. From the comparisons for photon beams, we can state that LegPy is able to calculate with reasonable accuracy 
%% FAM: agreement
(i.e., $\lesssim 5\%$)
the dose deposited by photons of up to 5 MeV in light media at depths of up to five mean free paths. This is particularly interesting for applications in medical physics
%%FAM: limitaciones del algoritmo.
as long as very high accuracy is not required.
The algorithm can also reproduce accurately the spectrum of absorbed energy in light elements as well
%%FAM: faltaba "as"
as
in typical scintillator materials used in gamma-ray spectrometry. The results of the angular distribution of escaping photons obtained with LegPy are also in
%%JR: definir agreement
%general
good agreement (around $1\%$)
with those from PENELOPE, but results for water at 5 MeV have significant deviations, indicating that the scattering of photons is more sensitive to the approximations employed in our algorithm. We found that the main limitation of our algorithm for the simulation of low-energy gamma rays is that it is not valid at energies close but higher than that of the K shell of heavy elements. We plan to implement soon a simplified model for the X-ray fluorescence in our algorithm to solve this problem.
%%JR: limitaciones del algoritmo
Even though pair production is also ignored, we found that this process is not very relevant for the dose deposited by photons of up to 5 MeV in heavy media.

Regarding the transportation of electrons, we have presented a very simple model that provides reasonable results of both the dose in depth and backscattering. Our results of the dose in depth in light elements deviate from those from PENELOPE by less than or around 10\% at the maximum of the depth in dose function. For heavy elements, deviations remain small up to a few MeV, but LegPy underestimates significantly the energy deposition at higher energies as a consequence of ignoring the Bremsstrahlung effect. In general, backscattering is properly simulated with LegPy
%%FAM
in heavy elements.
The most significant deviations with respect to PENELOPE were found for light elements, although they 
%%FAM
%are expected to
have
a minor effect in most practical cases because backscattering is only relevant for heavy elements.

LegPy has also been tested for the simulation of photons and electrons around the boundary between two media. The expected effects are properly observed and results are in good agreement with those obtained with PENELOPE for
%%FAM 
%both 
a 1 MeV photon beam ($\lesssim 3\%$), but the agreement is somewhat worse for electron beams due to the strong simplifications of our algorithm.
%

%%JR: reescribir.
%The several simplifications of our algorithm greatly improves the simulation speed. For example, the computing times needed to get the results of dose deposited by 5~MeV photons in lead shown in Fig. \ref{fig:B_Pb} were 130 times smaller for LegPy than for PENELOPE, in spite that LegPy simulations were run using a Python interpreter. At lower photon energies and for lighter elements, the number of interactions is reduced and the speed of both simulation codes becomes more similar. The increase in speed is also significant for electrons. For example, LegPy is faster than PENELOPE by a factor of 36 for 1.03 MeV electrons in aluminum (Fig. \ref{fig:Al_0.3-0.5-1.0}) if using a step length of 10\% of the CSDA range in LegPy. For a step length of 1\% of the CSDA, the speed of LegPy decreases approximately by a factor of 10, but it is still faster than PENELOPE.
LegPy was conceived to be used in a flexible and interactive way in a Google Colab or Jupyter notebook. In spite of using a Python interpreter, the several simplifications of our algorithm make in general the simulations rather fast compared to detailed simulation codes as PENELOPE, especially when the number of interactions is large (e.g., at high energy and heavy media). For example, the computing times needed to get the results of dose deposited by 5~MeV photons in lead shown in Fig. \ref{fig:B_Pb} were around 4 times smaller for LegPy than for PENELOPE when the tracking of secondary electrons is deactivated in both codes. The tracking of electrons is what demands the most CPU time, therefore our simplified model speeds up the simulation of electrons significantly. In the previous example, LegPy simulations including secondary electrons with an energy cut of 1~keV and a step length of 1 mm are faster by a factor of 80 than PENELOPE simulations with the same energy cut for electrons and tracking parameters of $C_1=C_2=0.05$ and $W_{\rm cc}=W_{\rm cr}=250$~eV. Obviously, the accuracy reached by PENELOPE is much higher.

In summary, an easy-to-use tool for fast simulations of low energy gamma-rays and electrons is available. LegPy aims to be useful for researchers that need simulations on simple geometries with reasonable accuracy
%%JR limitaciones. Ya indicado en el segundo párrafo de las conclusiones. Pero no está de más insistir en esta resumen
(i.e., around $10\%$)
but without the technical complexity of other MC codes. As the tool is very easy to use, it may also be useful for teaching purposes in undergraduate or postgraduate degree programs as well as for the training of experts in the field of medical applications of ionizing radiation.

\section*{Acknoledgements}
We thank Francesc Salvat for providing help and guidance for the use of PENELOPE program. We gratefully acknowledge financial support from the Spanish Research State Agency (AEI) through the grant PID2019-104114RB-C32. V. Moya also acknowledges the research grant CT19/23-INVM-109 funded by NextGenerationEU.

%see appendix~\ref{sec:sample:appendix}.

%\appendix

%\section{Sample Appendix Section}
%Lorem ipsum dolor sit amet, consectetur adipiscing elit, sed do eiusmod tempor section \ref{sec:sample1} incididunt ut labore et dolore magna aliqua. Ut enim ad minim veniam, quis nostrud exercitation ullamco laboris nisi ut aliquip ex ea commodo consequat. Duis aute irure dolor in reprehenderit in voluptate velit esse cillum dolore eu fugiat nulla pariatur. Excepteur sint occaecat cupidatat non proident, sunt in culpa qui officia deserunt mollit anim id est laborum.

% \bibliographystyle{elsarticle-num} 
% \bibliography{cas-refs}

\begin{thebibliography}{00}

\bibitem{egs4}
W.R. Nelson, H. Hirayama, D.W.O. Rogers, The EGS4 Code System, SLAC-265 (1985).

\bibitem{penelope}
J. Baró, J. Sempau, J.M. Fernández-Varea, F. Salvat., Nucl. Instrum. Meth. Phys. Res. Sect B 100, 31 (1995); doi: 10.1016/0168-583X(95)00349-5

\bibitem{geant4} 
S. Agostinelli, J. Allison, {\sl et al.},  Nucl. Instrum. Meth. Phys. Res. Sect A. 506, 250 (2003); doi: 10.1016/S0168-9002(03)01368-8

\bibitem{mcnp}
%T. Goorleya, M. Jamesb, {\sl et al.}, Nuclear Technology 180, 298 (2012); doi: 10.13182/NT11-135
%https://mcnp.lanl.gov/
J. A. Kulesza, T. R. Adams, {\sl et al.}, MCNP\textsuperscript{\textregistered} Code Version 6.3.0 Theory \& User Manual, LA-UR-22-30006, Los Alamos National Laboratory, Los Alamos, NM, USA (2022); doi: 10.2172/1889957

\bibitem{Salvat}
F. Salvat, PENELOPE-2018: A code system for Monte Carlo simulation of electron and photon transport, NEA/MBDAV/R(2019)1, OECD Nuclear Energy Agency, Boulogne-Billancourt; doi:10.1787/32da5043-en


\bibitem{legpy}
F. Arqueros, J. Rosado, V. Moya, LegPy (Version v1.2) [Computer software] (2022);   doi: 10.5281/zenodo.7882780

\bibitem{nist_ph}
M.J. Berger, J.H. Hubbell, S.M. Seltzer, {\sl et al.}, XCOM: Photon Cross Section Database (version 1.5) (2010). [Online] http://physics.nist.gov/xcom. National Institute of Standards and Technology, Gaithersburg, MD.

\bibitem{evans}
R. D. Evans, The Atomic Nucleus, McGraw-Hill (1955).
%reprinted (1982).R. E. Krieger, Malabar, FL.

\bibitem{nist_estar}
M. Berger, J. Coursey, M. Zucker, ESTAR, PSTAR, and ASTAR: Computer Programs for Calculating Stopping-Power and Range Tables for Electrons, Protons, and Helium Ions (version 1.21) (1999). [Online] http://physics.nist.gov/Star

\bibitem{theta_0}
G.R. Lynch and O.I. Dahl, Nucl. Instrum. Meth. Phys. Res. Sect B 58, 6 (1991); doi:10.1016/0168-583X(91)95671-Y

\bibitem{gcolab}
colab.research.google.com/

%\bibitem{amjphy}
%F. Arqueros, G. D. Montesinos, Am. J. Phys. 71, 38 (2003); doi: 10.1119/1.1509416

%\bibitem{coh_ef}
%D.E. Cullen, J.H. Hubbell, L. Kissel, EPDL97 The Evaluated Photon Data Library, 097 Version, Technical Report UCRL-50400, Lawrence Livermore National Laboratory, Livermore, California  (1997).

\bibitem{ansi-ans91}
American National Standard Institute. Gamma-ray attenuation coefficients and buildup factors for engineering materials. Report ANSI/ANS-6.4.3-1991. La Grange Park, Illinois: American Nuclear Society (1991).

\bibitem{hirayama}
H. Hirayama, J. Nucl. Sci. Technol. 32, 1201 (1995); doi: 10.1080/18811248.1995. 9731842

%H. Hirayama, Journal of Nuclear Science and Technology, Calculation of Gamma-ray Exposure Buildup Factors up to 40mfp using the EGS4 Monte Carlo Code with a Particle Splitting, Journal of Nuclear Science and Technology, 32:12, 1201-1207 (1995), DOI: 10.1080/18811248.1995.9731842

\bibitem{UEI} 
Uei-Tyng Lin, Shiang-Huei Jiang, Radiat. Phys. Chem. 48, 389 (1996); doi: 10.1016/0969-806X(95)00461-6

\bibitem{Manohara}
S.R. Manohara, S.M. Hanagodimath, L. Gerward,  J. Appl. Clin. Med. Phys. 12, 296 (2011); doi: 10.1120/jacmp.v12i4.3557. 
%Journal of Applied clinical Medical Physics (JACMP)

\bibitem{kadri}
O. Kadri, A. Alfuraih  Nucl. Sci. Tech. 30, 176 (2019); doi:10.1007/s41365-019-0701-4
%Photon energy absorption and exposure buildup factors for deep penetration in human tissues.
\bibitem{ansi_up_low}
L. Durani, "Update to ANSI/ANS-6.4.3-1991 for low-Z and compound materials and review of particle transport theory" (2009). UNLV Theses, Dissertations, Professional Papers, and Capstones. 43; 
doi: 10.34917/1363554

\bibitem{ansi_up_high}
L.P. Ruggieri, "Update to ANSI/ANS-643-1991 for high-Z materials and review of particle transport theory" (2008). UNLV Retrospective Theses Dissertations; doi:10.25669/xc1i-qbff

\bibitem{mcnp5}
X-5 Monte Carlo Team, MCNP — A General Monte Carlo N-Particle Transport Code, Version 5, Volume I: Overview and Theory, LA-UR-03-1987, Los Alamos National Laboratory (2003).

\bibitem{subbaiah}
K.V. Subbaiah, A. Natarajan, Nuclear Science and Engineering 96, 330 (1987); doi: 10.13182/NSE87-A16396
%Effect of Fluorescence in Deep Penetration of Gamma Rays. 

%\bibitem{subbaiah_2}
%Effect of Fluorescence, Bremsstrahlung, and Annihilation Radiation on the Spectra and Energy Deposition of Gamma Rays in Bulk Media
%K. V. Subbaiah, A. Natarajan, D. V. Gopinath & D. K. Trubey, Nuclear Science and Engineering 81, 172 (1982); doi: 10.13182/NSE82-A20084 

\bibitem{sandia_el_1}
G.J. Lockwood, G.H. Miller, L.E. Ruggles, J.A. Halbleib, Calorimetric measurement of electron energy deposition in extended media, SANDIA Tech. Rep. SAND79-0414, UC-34a, Albuquerque, NM (1987).

\bibitem{sandia_el_2}
G.J. Lockwood, G.H. Miller, L.E. Ruggles, J.A. Halbleib, Electron Energy and Charge Albedos — Calorimetric Measurement vs Monte Carlo Theory, Tech. Rep. SAND80-1968, UC-34a, Albuquerque, NM (1987).
%SANDIA REPORT SAND79-0414.UC-34a (1987)
%SAND80– 1968 UC–34a (1987)

%G.J. Lockwood, L.E. Ruggles, G.H. Miller, J.A. Halbleib, Calorimetric Measurement of Electron Energy Deposition in Extended Media—Theory vs Experiment, Sandia Laboratories Tech. Rep. SAND79-0414, Albuquerque, NM, 1980.

%\bibitem{Tabata_bck}
%T. Tabata,  Phys. Rev. 162, 336 (1967); doi:10.1103/PhysRev.162.336
%Backscattering of Electrons from 3.2 to 14 MeV,

\bibitem{Sempau}
J. Sempau, J.M. Fernandez-Varea, E. Acosta, F. Salvat, Nucl. Instrum. Meth. Phys. Res. Sect B 
207, 107 (2003); doi:10.1016/S0168-583X(03)00453-1 
%Experimental benchmarks of the Monte Carlo code PENELOPE

\bibitem{Kadri}
O. Kadri, V.N. Ivanchenko, F. Gharbi, A. Trabelsi, Nucl. Instrum. Meth. Phys. Res. Sect B 258, 381 (2007); doi:10.1016/j.nimb.2007.02.088
%GEANT4 simulation of electron energy deposition in extended media

\bibitem{P_Arce}
P. Arce,{\sl et al.}, Med. Phys. 48, 19 (2021); doi:10.1002/mp.14226
% Report on G4‐Med, a Geant4 benchmarking system for medical physics applications developed by the Geant4 Medical Simulation Benchmarking Group.

\bibitem{Salvat}
F. Salvat, PENELOPE-2018: A code system for Monte Carlo simulation of electron and photon transport, NEA/MBDAV/R(2019)1, OECD Nuclear Energy Agency, Boulogne-Billancourt; doi:10.1787/32da5043-en

\end{thebibliography}


\end{document}